\documentclass[reprint,
amsmath,amssymb,
aps,
prb,
floatfix,
]{revtex4-2}
\usepackage{ amssymb }

\usepackage{graphicx}

\usepackage{dcolumn}
\usepackage{color}
\usepackage{bm}
\usepackage{url} 
\usepackage{hyperref}
\hypersetup{
 colorlinks   = true,
 linkcolor=blue,
  citecolor    = blue,
  urlcolor=blue,
  } 

\usepackage{physics}

\usepackage[normalem]{ulem}
\usepackage[dvipsnames]{xcolor}
\newcommand{\anna}{\textcolor{black}}
\newcommand{\outstrike}[1]{}

\begin{document}

\title{Natural width of the superconducting transition in epitaxial TiN films}

\author{Elmira Baeva$^{1,2}$, Anna Kolbatova$^{1,*}$, Nadezhda Titova$^{1}$, Soham Saha$^{3}$, Alexandra Boltasseva$^{3}$, Simeon Bogdanov$^{4,5,6}$, Vladimir M.  Shalaev$^{3}$, Alexander Semenov$^{1}$, Gregory N. Goltsman$^{1,2}$, and Vadim Khrapai$^{2}$}

\affiliation{$^1$ Moscow Pedagogical State University, 119435 Moscow, Russia \\ $^2$ HSE University, 101000 Moscow, Russia \\$^3$ Birck Nanotechnology Center and Elmore Family School of Electrical and Computer Engineering, Purdue University, IN 47907 West Lafayette, USA \\$^4$ Department of Electrical and Computer Engineering, University of Illinois at Urbana-Champaign, IL 61801 Urbana, USA \\$^5$ Holonyak Micro and Nanotechnology Lab, University of Illinois at Urbana-Champaign, IL 61801 Urbana, USA \\ $^6$ Illinois Quantum Information Science and Technology Center, University of Illinois at Urbana-Champaign, IL 61801 Urbana, USA}


\begin{abstract}
We investigate the effect of various fluctuation mechanisms on the DC resistance in superconducting devices based on epitaxial titanium nitride (TiN) films. The samples we studied show a relatively steep resistive transition (RT), with a transition width $\Delta T/T_\mathrm{c} \sim 0.002-0.025$, depending on the film thickness (20 nm, 9 nm, and 5 nm) and device dimensions. This value is significantly broader than expected due to conventional superconducting fluctuations ($\Delta T/T_\mathrm{c} \ll 10^{-3}$). The shape and width of the RT can be perfectly described by the well-known effective medium theory, which allows us to understand the origin of the inhomogeneity in the superconducting properties of TiN films. We propose that this inhomogeneity can have both dynamic and static origins. The dynamic mechanism is associated with spontaneous fluctuations in electron temperature (T-fluctuations), while the static mechanism is due to a random spatial distribution of surface magnetic disorder (MD). Our analysis has revealed clear correlations between the transition width and material parameters as well as device size for both proposed mechanisms. While T-fluctuations may contribute significantly to the observed transition width, our findings suggest that the dominant contribution comes from the MD mechanism. Our results provide new insights into the microscopic origin of broadening of the superconducting transition and inhomogeneity in thin superconducting films.
\end{abstract}


\maketitle

\section{Introduction} \label{section_1}
It is well established that the transition from the normal state to the superconducting state occurs gradually, and is driven by dynamic fluctuations in the modulus and phase of the superconducting order parameter~\cite{Skocpol1975}. These so-called superconducting fluctuations manifest themselves in various macroscopic properties~\cite{Varlamov2018} and, most importantly, in DC resistance ($R$). Although the dynamics of these fluctuations are averaged over time and sample volume in resistance measurements, fluctuations in the order parameter modulus result in excess conductivity at temperatures above the critical temperature ($T_\mathrm{c}$)~\cite{Aslamasov1968, Maki1968}, while the phase fluctuations lead to a non-zero resistance below $T_\mathrm{c}$~\cite{TinkhamBook,Konig2021}. As a result, the superconducting resistive transition (RT) is broadened due to these fluctuations.

In general, theories that describe conventional superconducting (SC) fluctuations  assume the case of homogeneous superconductors.
However, many studies on structurally homogeneous superconductors have revealed an additional broadening of RT, in contrast to what would be expected due to SC fluctuations~\cite{Larkinlate2005, Benfatto2009, Caprara2011}. Additionally, a satisfactory agreement with experimental data was observed primarily for thin films with a high resistance in the normal state~\cite{Larkinlate2005}.
The observation of a broader RT in homogeneous superconductors is typically attributed to mesoscopic inhomogeneities, which can be related to either a spatial distribution of the order parameter~\cite{Sacepe2008, Hortensius2013, Venditti2019}, or the local superfluid stiffness~\cite{Benfatto2009, Mondal2011, Venditti2019}. In such cases, the resistive transition can be well described by a random-resistor network model, which is based on the so-called effective medium theory (EMT)~\cite{Caprara2011}. This model accounts for the transport through an inhomogeneous background, regardless of its microscopic origin.

In this study, we investigate the potential origin of the RT broadening observed in epitaxial TiN films. These films exhibit a high degree of crystallinity and exceptional electronic properties, implying a mean free path limited by surface scattering, a $T_\mathrm{c}$ value close to expected for bulk and a relatively steep transition to the superconducting state with a transition width $\Delta T/T_\mathrm{c} \sim 0.002-0.025$~\cite{Saveskul}. Despite the fact that the width of the RT observed in these high-quality films exceeds the value predicted by conventional SC fluctuation models, the shape of the RT dependencies can be accurately described using the EMT approach. 

In line with our previous experimental findings on epitaxial TiN films~\cite{Saveskul, first}, we propose two microscopic mechanisms for the RT broadening in homogeneous superconducting systems. 
Similar to conventional SC fluctuations, the first proposed mechanism has a dynamic nature and arises from spontaneous fluctuations in the electron temperature due to interactions with the phonon bath~\cite{first}. These fluctuations can cause local jumps between the normal and superconducting states of the system. We find that the predictions from the temperature fluctuation model ($T$-fluctuations) are in good agreement with the RT width at certain film thicknesses of TiN (9 nm and 20 nm), but these fluctuations alone are not sufficient to explain all of the observed results. The second mechanism involves static disorder, which is associated with fluctuations in the distribution of magnetic defects on the surface of TiN films ~\cite{Saveskul}.  The surface magnetic disorder can give rise to spatial inhomogeneity in $T_\mathrm{c}$ due to spin-flip scattering. In order to address this issue in a self-consistent manner, we have calculated the dispersion in $T_\mathrm{c}$ using the Ginzburg-Landau coherence length as a correlation length. Substituting this dispersion ($\delta T_\mathrm{c}$) in the standard EMT model provides good agreement with the experimental data for most of the TiN samples studied.

The article is organized as follows. Section~\ref{section_2} discusses in detail the samples and the experimental setup used in the study. Section~\ref{section_3} and section~\ref{section_4} analyze the experimental data from a conventional SC fluctuation model perspective. Section~\ref{section_5} examines the mechanisms underlying the observed broadening in the RT based on the experimental results. Section~\ref{section_6} discusses the overall findings.

\section{Samples and electrical resistance measurements}\label{section_2}

\begin{figure}[h!]
\centering
\includegraphics[scale=1]{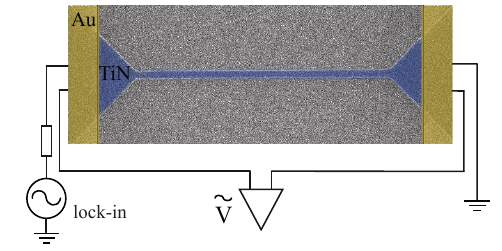}
    \caption{A false color image of a representative TiN sample (B6), obtained using a scanning electron microscope (SEM), is presented, along with a schematic diagram of  the electrical resistance measurement circuit.  \label{figure_num_1}}
\end{figure} 

The samples are made from epitaxial TiN films exhibiting monocrystalline quality and structural uniformity~\cite{Saveskul}. The TiN films with thicknesses of 5 nm, 9 nm, and 20 nm were grown on c-cut sapphire substrates using DC reactive magnetron sputtering from a 99.999\% pure Ti target at a temperature of 800$^{\circ}$C. The films were grown in an argon-nitrogen environment at a pressure of 5 mTorr with an Ar:N$_2$ flow ratio of 2:8\,sccm. Before lithographic processing, the films exhibited high critical temperatures ($T_\mathrm{c}$) and low normal-state sheet resistance ($R_\mathrm{s}$): $T_\mathrm{c} = 4.07$\,K, $R_\mathrm{s} = 60$\,$\Omega$/sq, $T_\mathrm{c} = 5.12$\,K, $R_\mathrm{s} = 9$\,$\Omega$/sq, and $T_\mathrm{c} = 5.6$\,K, $R_\mathrm{s} = 3.5$\,$\Omega$/sq for 5 nm, 9 nm, and 20 nm thick films, respectively. Here, the values of $R_\mathrm{s}$ are determined above the RT at 6\,K. The samples were patterned into bridges and meander structures using optical lithography, scanning electron beam lithography and plasma chemical etching. The samples, labeled as A1 to A3 and B6 to B10 and C1, as shown in table~\ref{T1}, have a two-contact bridge configuration, and in this case, the resistance was measured in a quasi four-probe (q4p) configuration. The samples, labeled as B1-B5, are Hall bars, and the resistance was measured in four-probe (4p) configurations. Figure~\ref{figure_num_1} shows a scanning electron microscope (SEM) image of a representative sample of TiN (B6) used in this study. Other configurations are presented in the supplementary material~\cite{Suppl_Data}. To characterize the quality of the samples, we estimate the sheet resistance after lithographic processing using the formula $R_\mathrm{s} = R_\mathrm{n}/N_\mathrm{s}$, where $R_\mathrm{n}$ represents the resistance in the normal state at 6\,K, $N_\mathrm{s} = L/w$ is equal to the number of squares in the sample, where $w$ and $L$ are the width and the length of the narrow part of samples.
Considering the contact resistance values of approximately 200 $\Omega$ and 3 $\Omega$ for two-terminal samples with thicknesses of 5 nm and 9 nm thick, as reported in the supplementary material~\cite{Suppl_Data}, the following values for the sheet resistance were obtained: $R_\mathrm{s}$ = 90 $\pm$ 10 $\Omega$/sq, 9.5 $\pm$ 0.5 $\Omega$/sq, and 3.5 $\pm$ 0.3 $\Omega$/sq for TiN with thicknesses of 5 nm, 9 nm and 20 nm, respectively.

The resistance measurement scheme shown in figure~\ref{figure_num_1} is performed using the standard lock-in technique. The bias current $I_\mathrm{ac}$ varies in the range from 6 to 500 nA at frequencies between 8 and 11 Hz. The AC voltage output signal is amplified using an SR560 preamplifier and measured using an SR830 lock-in. The experimental setup is designed with coaxial stainless steel lines and low-frequency RC filters. These filters consist of a 1 k$\Omega$ planar resistor and a 1 nF planar capacitor mounted on a common plate with the sample. The cutoff frequency of the filter is approximately 160 kHz. The bath temperature ($T_\mathrm{b}$) is measured using a calibrated RuO thermometer next to the sample. The critical temperature ($T_\mathrm{c}$) is determined as the point at which the sample lost half of its resistance, $R = R_\mathrm{n}/2$. Sample parameters such as thickness $d$, width $w$, length $L$,  resistance in the normal state $R_\mathrm{n}$ determined at 6\,K, critical temperature $T_\mathrm{c}$, circuit configuration, bias current $I_\mathrm{ac}$, and the coherence length $\xi_0$ are specified in table~\ref{T1}.

\begin{table}
\centering
\caption {\label{T1} Parameters of the studied TiN samples.}
\begin{tabular}{c c c c c c c c c c c}
\hline
\hline
    & $d$ & $L$    & $w$    & $N_\mathrm{s}$ & $R_\mathrm{n}$    & $T_\mathrm{c}$ & conf. & $I_\mathrm{ac}$   & $\xi_0$ \\ \hline
    & nm  & $\mu$m & $\mu$m &          & $\Omega$ & K     &  &     nA    & nm \\ \hline
A1  & 5   & 9.663  & 0.45   & 21.47    & 2.2k     & 4.073 & q4p & 6.8 & 13.7  \\
A2  &    & 2.8    & 0.11   & 25.46    & 2.3k     & 3.9  & q4p & 16 & \\
A3  &    & 50     & 0.37   & 129.6    & 10k      & 4    & q4p & 16 & \\
A4  &    & 1000     & 500   & 2    & 132      & 3.95    & 4p & 316 & \\
B1  & 9   & 1000   & 500    & 2        & 18       & 5.115 & 4p  &500    & 22 \\
B2  &    & 20     & 10     & 4        & 43       & 5.114 & 4p   &400    & \\
B3  &    & 20     & 3      & 8.67     & 92       & 5.112  & 4p  &200    &  \\
B4  &    & 20      & 1   & 22     & 237       & 5.106 & 4p & 100 & \\
B5  &    & 20      & 0.5   & 40     &430    & 5.1  & 4p & 200 & \\
B6 &    & 8      & 0.252  & 31.74    & 310      & 5.083 & q4p & 30 & \\
B7 &    & 8      & 0.15   & 53.33    & 522      & 5.068  & q4p & 15& \\
B8 &    & 8      & 0.083  & 86.38    & 995      & 5.048 & q4p & 15 & \\
B9 &    & 8      & 0.064  & 125      & 1.11k    & 5.036 & q4p & 9 & \\
B10 &    & 8      & 0.064  & 125      & 1.11k    & 5.064 & q4p & 20 & \\
C1 & 20   & 65  &0.28  &  240   & 800 & 5.25  & q4p & 7 & 24 \\
\hline
\hline
\end{tabular}
\end{table}

\section{Resistive transition and superconducting fluctuations} \label{section_3}

We begin our analysis by discussing the conventional mechanisms of superconducting fluctuations on the resistive transition~\cite{Larkinlate2005}. The effect of these fluctuations on the conductivity near $T_\mathrm{c}$ is typically discussed in terms of three contributions: the Aslamazov-Larkin (AL) corrections, which corresponds to a shunt effect due to fluctuating Cooper pairs~\cite{Aslamasov1968}; the anomalous Maki-Thompson (MT) term, generated by the coherent scattering of electrons forming a Cooper pair on impurities~\cite{Maki1968, Thompson1970}; and a term related to the fluctuation renormalization of density of states (DOS)~\cite{Ioffe1993}. These mechanisms collectively affect the overall conductivity, and consequently, the temperature dependence of the resistance. The latter can be described by taking into account the contributions from AL, MT, and DOS terms as follows: $R/R_\mathrm{n} = \left(1 + \rho_\mathrm{n}\left(\sigma_\mathrm{AL} + \sigma_\mathrm{MTreg + DOS} + \sigma_\mathrm{MTan} \right)\right)^{-1}$. Here $R/R_\mathrm{n}$ is the normalized resistance, $\rho_\mathrm{n} =  R_\mathrm{s} d$ is the normal-state resistivity. The description of the conductivity due to the superconducting fluctuations ($\sigma_\mathrm{AL}, \sigma_\mathrm{MTreg + DOS}, \sigma_\mathrm{MTan}$) is provided in the text below.

The effects of fluctuations increase as the dimension of a system decreases. The fluctuation conductivity due to the AL mechanism for the quasi two dimensional (2D) systems $(d < \xi(T) < w, L)$ and one dimensional (1D) systems $(d, w < \xi(T) < L)$ can be expressed as follows: 
\begin{subequations}\label{eq_AL} 
\begin{equation}
2D: \sigma_\mathrm{AL} = \frac{e^2}{16\hbar d}\epsilon_\mathrm{T}^{-1},
\label{eq_AL_2D}
\end{equation}
\begin{equation}
1D: \sigma_\mathrm{AL} = \frac{\pi e^2}{16\hbar d}\frac{\xi(0)}{w}\epsilon_\mathrm{T}^{-3/2}.
\label{eq_AL_1D}
\end{equation}
\end{subequations}
Here, $\xi(T) = \xi(0)/\sqrt{\epsilon_\mathrm{T}}$ is the coherence length, $\epsilon_\mathrm{T} = \ln(T/T_\mathrm{c0})$ is the reduced temperature and $T_\mathrm{c0}$ is the critical temperature renormalized by fluctuations~\cite{Larkinlate2005}. The fluctuation conductivity due to other mechanisms (MT and DOS) can be described by the following equations in 2D case~\cite{Larkinlate2005}:
\begin{equation}
\sigma_\mathrm{MTreg+DOS} = 0.691 \frac{e^2}{\hbar d}\ln\left( \epsilon_\mathrm{T}\right),
\label{eq_DOS}
\end{equation}
\begin{equation}
\sigma_\mathrm{MTan} = \frac{1}{8} \frac{e^2}{\hbar d \left( \epsilon_\mathrm{T}-\gamma_{\phi}\right)}\ln\left( \frac{\epsilon_\mathrm{T}}{\gamma_{\phi}}\right).
\label{eq_MT}
\end{equation}
 The anomalous MT contribution ($\sigma_\mathrm{MTan}$) contains a dimensionless phase-breaking parameter $\gamma_{\phi} = \pi \hbar/8k_\mathrm{B}T\tau_{\phi}$, where $\tau_{\phi}$ is the phase-breaking time, which usually depends on the temperature and the quality of the sample.

\begin{figure}[h!]
\centering
\includegraphics[scale=1]{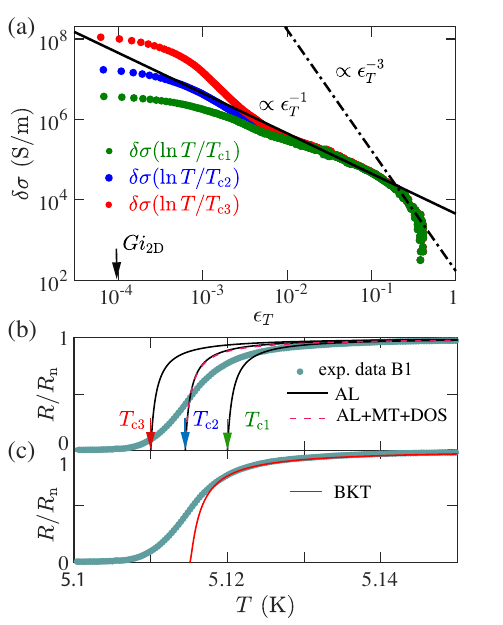}
\caption{(a) The dependence of the fluctuation conductivity, $\delta\sigma = \rho(T)^{-1}-\rho_\mathrm{n}^{-1}$, on the reduced temperature, $\epsilon_\mathrm{T} = \ln (T/T_\mathrm{c0})$, is shown for three different values of $T_\mathrm{c0}$ ($T_\mathrm{c1} = 5.12$\,K, $T_\mathrm{c2} = 5.1135$\,K and $T_\mathrm{c3} = 5.11$\,K are indicated by arrows in panel (b). The data are plotted for sample B1 on a log-log scale. The solid line corresponds to the 2D case of the AL model ($\propto \epsilon_\mathrm{T}^{-1}$), whereas the dash-dotted line represents the asymptotic $\epsilon_\mathrm{T}^{-3}$-behavior predicted for high temperatures, where the short-wavelength fluctuations become important~\cite{Johnson_1978,Hopf1991,Gauzzi1995}. (b-c) The normalized resistance, $R/R_\mathrm{n}$, plotted against $T$. The solid black lines shows the predictions of the AL model for all three values of $T_\mathrm{c0}$. The red dashed line represents the sum of the AL, MT, and DOS corrections to conductivity, calculated at $T_\mathrm{c2}$. It almost coincides with the AL model (black line). The solid red line represents the prediction of the BKT model.} \label{figure_num_2}
\end{figure}

First, we compare the experimental results for the dependence of $R(T)$ with the predictions of the AL model. However, as we will show below, a careful consideration of all known corrections to conductivity does not provide an adequate description of the $R(T)$ curve. Figure~\ref{figure_num_2}(a) shows the measured fluctuation conductivity, $\delta\sigma = \rho(T)^{-1} - \rho_\mathrm{n}^{-1}$, as a function of the reduced temperature, $\epsilon_\mathrm{T}$, for the sample B1, which represents a 2D fluctuation case. Dimensionality of the fluctuation regimes in the TiN samples was determined using the method described in the appendix. 
The fluctuation conductivity is calculated at three different values of $T_\mathrm{c0}$, indicated by arrows in figure~\ref{figure_num_2}(b). 
To compare the experimental data with equation~\eqref{eq_AL}, it is necessary to determine the value of $T_\mathrm{c0}$. 
However, this is a delicate point, as the value of $\delta\sigma$ diverges near the transition, and small changes in $T_\mathrm{c0}$ can significantly affect the character of the discrepancy. 
Figure~\ref{figure_num_2}(a) shows that, in the temperature range $\epsilon_\mathrm{T} \leq 5 \times 10^{-3}$, the slope $\delta\sigma$ is sensitive to the choice of $T_\mathrm{c0}$, while for $\epsilon_\mathrm{T} > 5 \times 10^{-3}$,  this is almost independent of $T_\mathrm{c0}$. We use $T_\mathrm{c0} = T_\mathrm{c2}$, where the measured fluctuation conductivity $\delta\sigma(T)$ is closest to the 2D asymptotic behavior ($\propto \epsilon_\mathrm{T}^{-1}$, as indicated by the blue symbols and the solid line in figure~\ref{figure_num_2}(a)). In the high temperature range $\epsilon_\mathrm{T} \geq 10^{-1}$, all experimental curves follow the modified theory of superconducting fluctuations, where the short-wavelength fluctuations are important~\cite{Johnson_1978,Hopf1991,Gauzzi1995} (as indicated by the $\epsilon_\mathrm{T}^{-3}$ behavior shown by the dash-dotted line in figure~\ref{figure_num_2}(a)). 
As can be seen in figure~\ref{figure_num_2}(b), the temperature dependence of $R/R_\mathrm{n}$ can be described by the AL expressions at temperatures significantly above $T_\mathrm{c0}$. Indeed, if we assume that $T_\mathrm{c0} = T_\mathrm{c1} > T_\mathrm{c2}$, the upper part of $R(T)$ can be approximated by the AL model to some extent, but the slope of $\delta\sigma(T)$ differs from the 2D AL model. If we choose $T_\mathrm{c0} = T_\mathrm{c2}$, then the slope of $\delta\sigma(T)$ will be closer to the AL slope, but this would still lead to insufficient description of the data even in the high-T region. A similar result is observed for $T_\mathrm{c0} = T_\mathrm{c3}$, as shown in figure~\ref{figure_num_2}. The same picture is also observed for other TiN samples (for details, see the supplementary material~\cite{Suppl_Data}). We also evaluate the MT and DOS contributions in comparison with the experimental data on TiN films. The previous results show that a small amount of surface magnetic disorder suppresses superconductivity at lower film thicknesses~\cite{Saveskul}. Using the Abrikosov-Gorkov theory~\cite{AG61}, we estimate a spin-flip scattering time $\tau_\mathrm{s} = 6$\,ps for the sample B1. This value of $\tau_\mathrm{s}$ is the shortest time among all the inelastic scattering times. These times also contribute to the total dephasing time $\tau_{\phi}$ in the MT term. Figure~\ref{figure_num_2}(b) illustrates the temperature dependence of resistance for B1 in comparison to the total effect from AL, MT, and DOS contributions (shown by the red dashed line). It can be noted that these fluctuations have a minimal impact on the $R(T)$ dependence. Therefore, it is important to consider the role of other factors in the broadening of the RT in epitaxial TiN films.

The deviation of the fluctuation conductivity from the AL value near $T_\mathrm{c0}$ is typically associated with the onset of critical fluctuations~\cite{Larkinlate2005}. The temperature range where critical fluctuations dominate is usually quantified using the Ginzburg-Levanyuk parameter. Typically, the Ginsburg-Levanyuk number $Gi$ is defined as (see equation~(2.87) in~\cite{Larkinlate2005}):
\begin{equation}
Gi = \frac{\Delta T}{T_c} = \left(\frac{7 \zeta(3)}{8 \pi^2} \frac{2\beta_\mathrm{D}}{N_\mathrm{0} k_\mathrm{B} T_\mathrm{c0}\xi(0)^\mathrm{D}}\right)^{2/\left(4-\mathrm{D}\right)},
\label{eq_Gi}
\end{equation}
where $N_\mathrm{0}$ is the electronic density of states at the Fermi level, $\mathrm{D}$ is the sample dimensionality, and $\zeta(3) = 1.206$. The constants $\beta_\mathrm{D}$ are defined as: $\beta_2 = 1/(4\pi d)$ for $\mathrm{D} = 2$ and $\beta_1 = 1/(4 d w)$ for $\mathrm{D} = 1$. Replacing $N_\mathrm{0} = (e^2\rho_\mathrm{n} \mathcal{D})^{-1}$ and $\xi(0)^2 = \pi\hbar \mathcal{D}/(8k_\mathrm{B}T_\mathrm{c})$ (where $\mathcal{D}$ is the diffusion coefficient) into equation~\eqref{eq_Gi} gives $Gi_\mathrm{2D} = e^2R_\mathrm{s}/23\hbar$ in the 2D case and $Gi_\mathrm{1D} = \left(e^2 R_\mathrm{s}\xi(0)/7.35\hbar w\right)^{2/3}$ for 1D.
The onset of critical fluctuations can be estimated by assuming that the reduced temperature, $\epsilon_\mathrm{T}$, is equal to the Ginzburg-Levanyuk number, $Gi$, which is approximately $10^{-4}$ for a 2D system. Note that the onset is expected to occur at $R/R_\mathrm{n} = 16/39 \approx 0.4$ for both 2D and 1D fluctuation cases. As shown in figure~\ref{figure_num_2}, the data is not well described by the AL model for values of $R$ less than $0.9R_\mathrm{n}$.

\section{BKT transition} \label{section_4}

Next, we will discuss the contribution of phase fluctuations of the order parameter to finite resistance on the low temperature side of the resistive transition. The superconducting transition in 2D samples is expected to exhibit the Berezinskii-Kosterlitz-Thouless (BKT) transition, in which a finite superfluid density is destroyed at the BKT temperature, $T_\mathrm{BKT}$, due to proliferation of vortex-antivortex pairs. This effect occurs when a logarithmic interaction between vortices dominates, i.e., the sample is narrower than the Pearl length $2\lambda_\mathrm{L}(T)^2/d$, where $\lambda_\mathrm{L}(T) = \lambda_\mathrm{L}(0)/\sqrt{1 - (T/T_\mathrm{c})^4}$ and $\lambda_\mathrm{L}(0) = \sqrt{m/(\mu_\mathrm{0}nq^2)} \approx 24$\,nm is the London penetration depth at absolute zero. This corresponds to the regime of 2D fluctuations~\cite{goldman1984percolation}. 

At temperatures close to $T_\mathrm{c0}$, but above $T_\mathrm{BKT}$, thermal fluctuations lead to the dissociation of bound vortices, which become free and move under the influence of a bias current, causing finite resistance. In clean samples, the fluctuation regime above $T_\mathrm{BKT}$ is usually dominated by AL fluctuations, while BKT fluctuations are limited to a narrow range near $T_\mathrm{BKT}$. Within this range, the resistance decreases exponentially. To account for this behavior in epitaxial TiN films, we use an interpolation formula that combines the resistance from the vortices in the superconducting state with the contribution from the AL mechanism above $T_\mathrm{c0}$~\cite{Halperin1979,Konig2015}:
\begin{equation}
\frac{R}{R_n}=\left[1+\left(\frac{2}{A}\sinh\frac{b}{\sqrt{t}}\right)^2\right]^{-1}. 
\label{bkt1}
\end{equation}
Here $t = (T - T_\mathrm{BKT})/T_\mathrm{BKT}$, and $A$ and $b$ are dimensionless fitting parameters. It should be noted that $T_\mathrm{BKT}$ is shifted relative to $T_\mathrm{c0}$ by a factor of $T_\mathrm{BKT} = T_\mathrm{c0}\left(1 - 4 Gi_\mathrm{2D}\right)$~\cite{Larkinlate2005}. 
As shown in figure~\ref{figure_num_2}(c), the normalized resistance measured for sample B1 corresponds to equation~\eqref{bkt1} with the best-fit values $T_\mathrm{c0} = 5.117$\,K, $Gi_\mathrm{2D} \approx 10^{-4}$, $T_\mathrm{BKT} \approx 5.115$\,K, $A = 1.3$, and $b = 0.011$. The value of $b$ correlates with a small distance between the mean field critical temperature and the BKT temperature: $t_\mathrm{c} = (T_\mathrm{c0} - T_\mathrm{BKT})/T_\mathrm{BKT}\approx 4\times10^{-4}$~\cite{Benfatto2009}. This indicates that the BKT transition may be indistinguishable from $T_\mathrm{c0}$, since it is limited by the temperature range $t \ll t_\mathrm{c}$, while above it, it restores the AL paraconductivity. Note that these measurements were performed in high-quality films with $R_\mathrm{s} \ll \hbar/e^2$, that is, the films are far from the superconductor-insulator transition~\cite{Gantmakher_2010}. In this case, the effect of disorder on  $T_\mathrm{c0}$ and $T_\mathrm{BKT}$, as well as on electronic properties, is considered to be negligible.

As shown in other studies~\cite{Benfatto2009,Mondal2011,Baity2016}, the $R(T)$ dependence in figure~\ref{figure_num_2}(c) cannot be explained without considering the intrinsic inhomogeneity of the electronic properties in a superconductor.
This inhomogeneity can be defined by the spatial variation of $T_\mathrm{c}$ or the superfluid density, and can be modeled using the effective medium theory (EMT). In the following section, we will discuss the experimental $R(T)$ dependencies in the EMT approximation and explore the possible microscopic origins of inhomogeneities in epitaxial TiN films.

\section{Microscopic description}\label{section_5} 
\subsection{The effective medium theory} \label{subection_5_1} 
\begin{figure}[h!]
\centering
\includegraphics[scale=1]{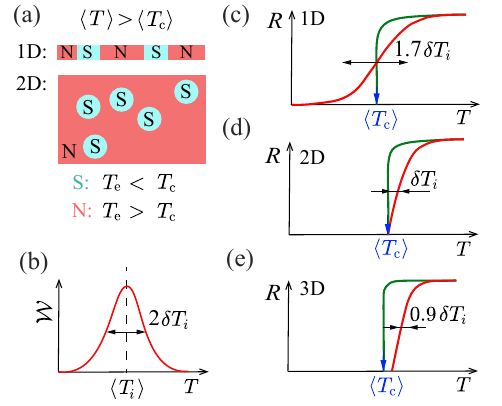}
\caption{Schematic representation of the resistive transition in the EMT method. (a) A sketch of various configurations (2D or 1D) of superconducting (S) samples with normal (N) regions that appear randomly at temperatures above the average critical temperature, $\langle T\rangle > \langle T_\mathrm{c}\rangle$. Normal patches occur when the local electron temperature of the resistors is above their critical temperature, $T_\mathrm{e} > T_\mathrm{c}$. The superconducting medium corresponds to the condition $T_\mathrm{e} < T_\mathrm{c}$. (b) The Gaussian distribution function $\mathcal{W}(T)$, which depends on the bath temperature $T$, is shown. $\delta T_\mathrm{i} \equiv \left[\delta T_\mathrm{c}, \delta T_\mathrm{e}\right]$ represents either the rms fluctuations of the critical temperature or the electron temperature. (c-e) A schematic representation of the temperature dependencies of resistivity for a single RRN resistor $\rho_\mathrm{0}(T)$ (green line) and the effective medium resistivity $\rho_\mathrm{EM}(T)$ (red line), for 1D (c), 2D (d), and 3D (e) cases.}\label{figure_num_3}
\end{figure}

To account for the inhomogeneity, we will discuss a model based on the effective medium theory (EMT)~\cite{Caprara2011}. This model considers the transition from a metal to a superconductor, where percolation through an inhomogeneous background dominates. At bath temperatures below the mean field critical temperature, the inhomogeneous background can be represented as a network of normal regions embedded in a superconducting matrix, which manifests itself mesoscopic length scales. See figure~\ref{figure_num_3} for a schematic illustration of this model. The sample is modeled as a random resistor network (RRN). This network consists of random resistors located on bonds, and is characterized by the effective medium resistivity $\rho_\mathrm{EM}$. We will use this model to account for both static and dynamic inhomogeneity in the system. In the first case, inhomogeneity is determined by local variations in the critical temperature ($\Delta T_c$). In the second case, it is related to fluctuations in an electron temperature ($\Delta T_e$). The critical temperature $T_c$ and the electron temperature $T_e$ can be assigned to each resistor in the RRN according to a given probability distribution. 
Following the analysis presented in the reference~\cite{Caprara2011}, we assume that the probability distribution of the critical temperature, denoted by $\mathcal{W}(T_c)$, follows a Gaussian shape, as shown in figure~\ref{figure_num_3}(b).
\begin{equation}
\mathcal{W}(T_\mathrm{c}) = \frac{1}{\sqrt{2\pi}\,\anna{\langle\Delta T_\mathrm{c}^2\rangle^{1/2}}}\exp\left(-\frac{\anna{\Delta T_\mathrm{c}^2}}{2\,\anna{\langle\Delta T_\mathrm{c}^2\rangle}}\right), \label{eq8} 
\end{equation}
where $\Delta T_\mathrm{c} = T_\mathrm{c} - \langle T_\mathrm{c}\rangle$ is the fluctuation of the critical temperature, the symbol $\langle \ldots\rangle$ represents the average over time and a correlation volume. By analogy, we can introduce the probability distribution of the fluctuation of the electron temperature, $\mathcal{W}(T_\mathrm{e})$, where $\Delta T_\mathrm{e} = T_\mathrm{e} - \langle T_\mathrm{e}\rangle$ with $\langle T_\mathrm{e}\rangle \equiv T$. The dispersion of the critical temperature and the electron temperature are $\delta T_\mathrm{c} \equiv \sqrt{\langle \Delta T_\mathrm{c}^2\rangle}$ and  $\delta T_\mathrm{e} \equiv \sqrt{\langle \Delta T_\mathrm{e}^2\rangle}$, respectively.

When the bath temperature $T$ decreases, some resistors enter a superconducting state ($\rho_\mathrm{i} = 0$) as soon as the condition $T_\mathrm{e} \leq T_\mathrm{c}$ is met. Other resistors remain in their normal state ($\rho_\mathrm{i} = \rho_\mathrm{0}$) at $T_\mathrm{e} > T_\mathrm{c}$. Regardless of the mechanism of the RT broadening, the effective medium resistivity $\rho_\mathrm{EM}$ obeys the equation:
\begin{equation}
\sum\limits_{i} \omega_\mathrm{i} \frac{\rho_\mathrm{EM}-\rho_\mathrm{i}}{\rho_\mathrm{EM} + \eta \rho_\mathrm{i}} = 0, \label{emt}
\end{equation}
where $\omega_i$ is the probability of the resistivity $\rho_i$ occurring and the parameter $\eta$ has values 2, 1, and 0 for 3D, 2D, and 1D cases, respectively. Note that equation~\eqref{emt} only applies when $\rho_\mathrm{EM} > 0$, which is true for all temperatures in 1D. However, in 2D and 3D this condition only holds for temperatures above a certain threshold, below which percolation along an infinite superconducting cluster occurs. As shown in figure~\ref{figure_num_3}(c-e), the center of RT coincides with the average value of $ T_c$ in 1D, whereas in 2D and 3D it shifts towards higher temperatures. The numerical coefficients here represent the relationship between the dispersion of fluctuations in $T_\mathrm{c}$ or $T_\mathrm{e}$ and the EMT transition width, as well as the sample dimension. The EMT transition width is defined numerically as a temperature range between 0.2 and 0.8 $R_\mathrm{n}$. The equation~\eqref{emt} is solved for 1D and 2D cases relevant to our experiment. In 1D case, the solution is straightforward:
\begin{equation}
\rho_\mathrm{EM} = \int^\infty_\mathrm{0} \rho_\mathrm{0}(T)\mathcal{W}(T)\dd T. \label{eq_emt_1d}
\end{equation}

In 2D, the EMT resistivity at a given temperature $T$ can be obtained by solving the equation numerically:
\begin{equation}
\int^\infty_{\langle T_\mathrm{c}\rangle} \frac{2\mathcal{W}(T)\dd T}{\rho_\mathrm{EM}/\rho_\mathrm{0}(T)+1} = 1.\label{eq11}
\end{equation}
In the above equations~\eqref{eq_emt_1d}- \eqref{eq11}, we approximate the resistivity $\rho_\mathrm{0}(T)$, the resistivity of a single resistor without spatial inhomogeneities, using the AL theory: $\rho_\mathrm{0}(T) = \left(1 + \rho_\mathrm{n} \sigma_\mathrm{AL}\right)^{-1}$. The width of the $\rho_\mathrm{0}(T)$ curve is determined by the product $Gi\,T_\mathrm{c}$, which is smaller than the rms fluctuations $\delta T_c$ or $\delta T_e$ in the EMT theory. This is shown schematically by green lines for $\rho_\mathrm{0}(T)$ and red lines for $\rho_\mathrm{EM}$ in panels (c)-(e) of figure~\ref{figure_num_3}.

\begin{figure}[h!]
\centering
\includegraphics[scale=1]{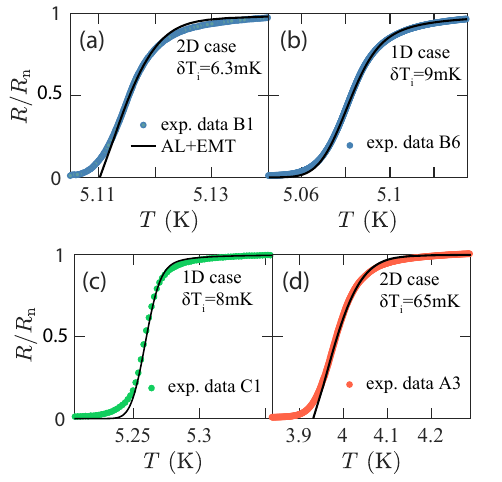}
   \caption{The temperature dependence of the normalized resistance $R/R_\mathrm{n}$ in comparison with the EMT approximation (equations~\eqref{eq_emt_1d}-\eqref{eq11}). The experimental data, represented by symbols, were obtained for four representative TiN samples (A3, B1, B6, C1). Considering the mechanisms of inhomogeneity in epitaxial TiN films discussed in the next sections, we applied the 1D model ($d, w < L_\mathrm{cor}$) for fitting samples B6 and C1 and the 2D model ($d < L_\mathrm{cor} < w$) for samples B1 and A3. Here, $L_\mathrm{cor}$ represents the correlation length, which chosen for describing static inhomogeneity (see section~\ref{subection_5_3}).}\label{figure_num_4}
\end{figure} 
 
Figure~\ref{figure_num_4} presents the experimental temperature dependencies of the normalized resistance $R/R_\mathrm{n}$ (shown by symbols) and the results obtained with the EMT approximation (shown by black solid lines). The experimental data are presented for four representative samples (B1, B6, C1 and A3, as described in table~\ref{T1}), the rest of the data can be found in the supplementary material~\cite{Suppl_Data}. Considering the potential mechanisms of inhomogeneity in epitaxial TiN films, which are discussed in further sections, the 1D model ($d, w < L_\mathrm{cor}$) was applied for fitting samples B6 and C1, while the 2D model ($d < L_\mathrm{cor} < w$) was used for samples B1 and A3. Here, $L_\mathrm{cor}$ represents the correlation length, which is the scale of the inhomogeneity. The best fit values of the temperature dispersion $\delta T_\mathrm{i}$, which are 6.3\,mK, 9\,mK, 8\,mK, and 65\,mK for B1, B6, C1, and A3 devices, respectively, perfectly describe the RT width of all devices. Below, we discuss  microscopic mechanisms that can lead to finite RT widths in epitaxial TiN films and the choice of the effective device dimension for the EMT approximation.

\subsection{ Dynamic inhomogeneity: T-fluctuations}\label{subection_5_2}
Next, we will discuss the possibility of the RT broadening due to slow dynamic fluctuations in electron temperature. As shown in the separate manuscript~\cite{first}, the agreement between the T-fluctuations model and noise measurements prompts us to consider the possible effect of T-fluctuations on DC resistance in homogeneous superconductors.

Microscopically, during T-fluctuations, the electronic system in a large sample cools or heats up spontaneously and randomly due to stochastic energy exchange with a thermal bath. This leads to a change in the temperature landscape over time and space. A finite correlation time determines the finite correlation length. As shown in reference~\cite{first}, this length is controlled by the electron-phonon scattering in TiN films. Therefore, it is defined as $L_\mathrm{cor} \equiv L_\mathrm{eph} = \sqrt{\mathcal{D}\tau_\mathrm{eph}}$, where $\mathcal{D}$ is the diffusion coefficient and $\tau_\mathrm{eph}$ is the electron-phonon relaxation time. This means that the fluctuations in temperature are completely independent at two points separated by a distance greater than $L_\mathrm{eph}$. More specifically, $\langle \delta T_\mathrm{i}(r_\mathrm{i})\delta T_\mathrm{j}(r_\mathrm{j})\rangle \propto \exp(-\abs{r_\mathrm{i} - r_\mathrm{j}}/L_\mathrm{eph})$, which allows us to determine the correlation volume $\mathcal{V}_\mathrm{c}$ of $T$-fluctuations. This correlation volume depends on the effective size of the device: $\mathcal{V}_\mathrm{c} = L_\mathrm{eph}^3$ in the 3D case ($d, w, L > L_\mathrm{eph}$), $\mathcal{V}_\mathrm{c} = dL_\mathrm{eph}^2$ in the 2D case ($w, L > L_\mathrm{eph} > d$), and $\mathcal{V}_\mathrm{c} = wd L_\mathrm{eph}$ in the 1D case ($d, w < L_\mathrm{eph} < L$). The individual T-fluctuation within a correlation volume, $\delta T_i^2\equiv\langle \delta T_\mathrm{e}^2\rangle = k_\mathrm{B}T^2/C_\mathrm{e}\mathcal{V}_\mathrm{c}$, remains finite even at $\mathcal{V} \rightarrow \infty$. Here, $C_\mathrm{e}$ is the electron heat capacity. Considering the fact that the length scale underlying the $T$-fluctuations is $L_\mathrm{eph} \sim 1.5$\,$\mu$m at $T_\mathrm{c}$ (for details, see ref.~\cite{first}), which is much larger than the GL coherence length, as well as all other transport length scales, it is natural to treat the fluctuating temperature landscape as a spatial inhomogeneity of $T_\mathrm{c}$~\cite{Caprara2011}.

In this section, we compare the values of $\delta T_\mathrm{i}$ obtained as a result of approximation using the EMT model with the T-fluctuations within one correlation volume, $\delta T_\mathrm{e}$. The electron heat capacity considered is $C_\mathrm{e} = \pi^2 k_\mathrm{B}^2T_\mathrm{e} N_\mathrm{0}/3$ for a free electron gas~\cite{Kittel}, where $N_\mathrm{0} = 1/(e^2\rho_n \mathcal{D})\simeq 75\pm 10$\,eV$^{-1}$nm$^{-3}$ is the estimated DOS at Fermi level, $\delta T_\mathrm{e}$ is expected to be 3.5$\pm$0.3\,mK, 8.5$\pm$0.3\,mK, 5.3$\pm$ 0.3\,mK, and 8.5$\pm$1.0\,mK, respectively, for B1, B6, C1, and A3 (the samples presented in figure~\ref{figure_num_4}). These estimates were obtained by considering different fluctuation regimes for the samples: B1 is in the 2D regime of T-fluctuations ($d < L_\mathrm{eph} < w, L$), and B6, C1, and A3 are in the 1D regime ($d, w < L_\mathrm{eph} < L$). 

A more or less satisfactory agreement between the experimental data and the T-fluctuation model is observed for samples with a thickness of 9\,nm and 20\,nm (B1 and B6, respectively, and C1). In these samples, the relative contribution of the temperature fluctuations to the RT width is approximately $\delta T_\mathrm{e} \approx 0.5\delta T_i$. This result indicates that the temperature fluctuations taken into account with the EMT approximation better describe the resistive transition compared to the traditional theory of SC fluctuations. However, for the device with a thickness of 5\,nm (A3), $\delta T_\mathrm{e} \ll \delta T_\mathrm{i}$, and it is obviously insufficient to to explain the observed RT width. This suggests that there may be another microscopic mechanism contributing to the broadening of the RT in TiN epitaxial films. We propose that randomly distributed surface magnetic disorder may play a significant role in this process. In the next section, we will provide evidence to support this hypothesis.

\subsection{ Static inhomogeneity: surface magnetic disorder} \label{subection_5_3}
In this section, we discuss the microscopic mechanism of spatial inhomogeneity of $T_\mathrm{c}$, caused by surface magnetic disorder (MD). As possible sources of local magnetic pair breaking, one can consider oxidized surface or interface areas where the variation in oxygen content leads to a number of unpaired electron orbitals.
Signatures of surface MD have been experimentally observed in nominally nonmagnetic superconductors~\cite{Proslier2008, Sendelbach2008, deGraaf2020, Fang2020, tamir2021direct, Kuzmiak2022}, and its presence is attributed to dangling bonds on the surface native oxide~\cite{deGraaf2020,tamir2021direct}. 

In our previous work~\cite{Saveskul}, we suggested that the variation of $T_\mathrm{c}$ with film thickness in nominally identical epitaxial TiN films can also be attributed to the presence of the surface MD residing in the surface oxide layer. The well-known detrimental effect on $T_\mathrm{c}$, due to pair-breaking spin-flip scattering, can be described with the Abrikosov-Gorkov (AG) equation~\cite{AG61} $\ln(T_\mathrm{c}^0/T_\mathrm{c}) = \psi(1/2 + x) - \psi(1/2)$, where $T_\mathrm{c}^0$ is the critical temperature in the absence of the magnetic disorder, $x = \hbar/(2\pi k_\mathrm{B}T_\mathrm{c}\tau_\mathrm{s})$ is the normalized spin-flip rate, and $\psi(x)$ is the digamma function. For small enough spin-flip scattering rate $\tau_\mathrm{s}^{-1}$, which moderates decrease in $T_\mathrm{c}$~\cite{Saveskul}, the AG equation can be approximated by a linear expression:
\begin{equation}
   T_\mathrm{c}^0 - T_\mathrm{c} = \frac{\pi\hbar}{4k_\mathrm{B}\tau_\mathrm{s}} = \frac{\pi\hbar v_\mathrm{F} a^2}{4k_\mathrm{B} d}\cdot n_\mathrm{s},
   \label{eq_ag_linearized}
\end{equation}
where the spin-flip rate have been expressed in terms of the surface density of magnetic impurities $n_\mathrm{s}$, the bulk critical temperature $T_\mathrm{c}^0$, the Fermi velocity $v_\mathrm{F}$, and the lattice constant of TiN $a\approx0.4\,$nm. In agreement with previous work (see table~\ref{T1} in~\cite{Saveskul}), the observed decrease in $T_\mathrm{c}$ as $d$ decreases is consistent with a characteristic density of magnetic defects $n_\mathrm{s} \sim  10^{16}\mathrm{m}^{-2}$.

Equation~\eqref{eq_ag_linearized} establishes a relationship between the average $T_\mathrm{c}$ and the average density of magnetic defects, $n_\mathrm{s}$. Similarly, local variations in $n_\mathrm{s}$ are expected to cause variations in the critical temperature, $\Delta T_\mathrm{c}$. This suggests that a stochastic spatial distribution of magnetic defects can lead to a built-in inhomogeneity in epitaxial TiN films and result in the observed broadening of RT~\cite{Caprara2011,Benfatto2009}. In this work, we assume a completely random distribution of magnetic defects, which is different from the case of magnetic superconductors~\cite{Koshelev2020}. This means that the number of impurities $N_\mathrm{m}$ in a given area $A$ of the film fluctuates as $\Delta N_\mathrm{m}(A) = \sqrt{N_\mathrm{m}(A)} = \sqrt{An_\mathrm{s}}$, which corresponds to a fluctuation in defect density $\Delta n_\mathrm{s}(A) = \sqrt{n_\mathrm{s}/A}$. In the absence of spatial correlations in the distribution of magnetic defects, it is important to choose a typical spatial scale that determines the value of $A$ (the correlation area). We treat the problem in a self-consistent manner, assuming that the relevant scale is given by the GL-like correlation length $\xi = \sqrt{\pi\hbar \mathcal{D}/8k_\mathrm{B}\Delta T_\mathrm{c}}$, which itself depends on the random fluctuation in the critical temperature $\Delta T_\mathrm{c}$ via equation~\eqref{eq_ag_linearized}. Depending on the ratio of $\xi$ and $w$, two regimes can be identified: a 2D regime ($w > \xi$) and a 1D regime ($w < \xi$), which correspond to the cases where $A = \xi^2$ and $A = w\xi$, respectively. Therefore, the rms fluctuation of $T_\mathrm{c}$ can be expressed as:
\begin{equation}
2D:\delta T_\mathrm{c} = \frac{2(T_\mathrm{c}^0 - T_\mathrm{c})v_\mathrm{F} a^2}{d\mathcal{D}},
\label{eq_mag2D}
\end{equation}

\begin{equation}
1D:\delta T_\mathrm{c} = \left (\frac{\pi \hbar (T_\mathrm{c}^0 - T_\mathrm{c})^2v_\mathrm{F}^2 a^4}{2 k_\mathrm{B}d^2w^2\mathcal{D}}\right)^{1/3}.
\label{eq_mag1D}
\end{equation}
The crossover from 2D to 1D regime occurs at a specific width of the sample $w_{MD} = \sqrt{\pi\hbar d \mathcal{D}^2/16k_B(T_\mathrm{c}^0 - T_\mathrm{c}) v_\mathrm{F}a^2}$. We find $w_{MD} \approx$ 0.07 $\mu$m, 0.5 $\mu$m, and 1.1 $\mu$m for films with thicknesses of 5 nm, 9 nm, and 20 nm, respectively. By comparing the device width to $w_{MD}$, the RT data in figure~\ref{figure_num_4} can be fitted using either equation~\eqref{eq_emt_1d} or equation~\eqref{eq11}, corresponding to the 1D and 2D EMT approximations. The results for the other samples are presented in the supplementary material~\cite{Suppl_Data}. 

\begin{figure}[t!]
\centering
\includegraphics[scale=1]{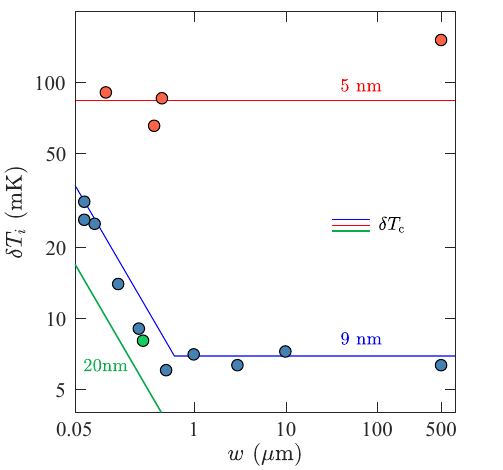}
    \caption{The RT width, denoted as $\delta T_\mathrm{i}$, is plotted as a function of the sample width, $w$. Symbols correspond to the best EMT fits of the experimental data, while solid lines represent the rms fluctuation of the critical temperature, $\delta T_\mathrm{c}$, from the MD model. Here, the same fitting parameters of $T_\mathrm{c}^0 = 5.6$\,K and $v_\mathrm{F} \approx 4\times 10^5\,$m/s are used to fit the data for all thicknesses.}   \label{figure_num_5}
\end{figure} 

Further, we plot the fitting parameter $\delta T_i$ for all devices under study (shown by symbols) compared to the MD model predictions (shown by solid lines) in figure~\ref{figure_num_5}. The predictions of the MD model were obtained for all three film thicknesses, considering the same fitting parameters $T_\mathrm{c}^0 = 5.6$\,K and $v_\mathrm{F} \approx 4\times 10^5\,$m/s. One can see that the MD model not only provides quantitatively similar results for samples with different thicknesses, but also accurately describes the crossover point from 2D to 1D behavior in the experimental data for samples fabricated from 9\,nm thick film. Therefore, we believe that magnetic disorder is the dominant factor responsible for the broadening of RT in epitaxial TiN films.

\section{Discussion} \label{section_6} 
We analyze the resistive transition in thin epitaxial TiN films, which exhibit a relatively steep transition to the superconducting state. As shown in sections~\ref{section_3}-\ref{section_4}, the width of the RT in these high quality films differs from predictions based on the conventional theory of superconducting fluctuations. Meanwhile, as shown in section~\ref{subection_5_1}, the shape of RT dependencies is almost perfectly described by the EMT approach, which takes into account inhomogeneities, regardless of their microscopic origin. In this study, we describe the RT dependencies microscopically by assuming either a dynamic or static origin for the RT broadening. The dynamic origin is attributed to spontaneous fluctuations of the electron temperature at fixed $T_\mathrm{c}$ (section~\ref{subection_5_2}), while the static origin corresponds to the in-built spatial fluctuations in $T_c$ (section~\ref{subection_5_3}). These fluctuation mechanisms can be conceptually described using the EMT approach, although they differ in terms of their dispersion ($\delta T_\mathrm{e}$ or $\delta T_\mathrm{c}$) and crossover points between 1D and 2D regimes. 

In this section, we present a summary of our findings for all the samples studied. We extract experimental values for the RT width, $\delta T_\mathrm{i}$, using the EMT approach. We assess the relative contribution of T-fluctuations and magnetic disorder to the broadening of the RT. Note that the effective dimensionality of a device in our data analysis with EMT is consistent with the proposed microscopic models. As discussed in section~\ref{subection_5_2}, we observe a significant contribution from temperature fluctuations for samples with thicknesses of 9 nm and 20 nm ($\delta T_\mathrm{e} \sim (0.3-0.6)\delta T_\mathrm{i}$), and a much smaller contribution for 5 nm devices ($\delta T_\mathrm{e}  < 0.1\delta T_\mathrm{i}$). However, the values of $\delta T_\mathrm{i}$ for 5 nm devices, as well as those for other thicknesses, can be well described by fluctuations of $T_\mathrm{c}$ due to magnetic disorder, as shown in figure~\ref{figure_num_5} in section~\ref{subection_5_3}). The latter suggests that the broadening of the RT in TiN films is likely related to the spatial distribution of $T_\mathrm{c}$.

As discussed in section~\ref{section_5}, two microscopic mechanisms of the RT broadening can be distinguished based on the dependence of the RT width $\delta T_\mathrm{i}$ on sample dimensions and the diffusion coefficient. In the 2D regime, $\delta T_\mathrm{e} \sim (d\mathcal{D})^{-1/2}$ for the T-fluctuations model and $\delta T_\mathrm{c} \sim (d^2\mathcal{D})^{-1}$ for the MD model. Comparison of $\delta T_\mathrm{i}$ for wide samples made from 5 nm and 9 nm TiN is inconsistent with the T-fluctuations and closely matches the MD model estimate. In 1D, the broadening of the RT also depends on device width: $\delta T_\mathrm{e} \sim d^{-1/2}w^{-1/2}\mathcal{D}^{-1/4}$ and $\delta T_\mathrm{c} \sim d^{-4/3}w^{-2/3}\mathcal{D}^{-1/3}$. Although the exponents of $w$ do not differ significantly, the data for the 9 nm films in the 1D regime (see figure~\ref{figure_num_5}) is again closer to the MD model prediction. Furthermore, a crossover between the 1D and 2D fluctuation regimes occurs at different widths in these models: $w \approx L_\mathrm{eph}\sim \mathcal{D}^{-1/2}$ for the T-fluctuations and $w \approx w_\mathrm{MD}\sim d\mathcal{D}$ for the MD model. The data in figure~\ref{figure_num_5} seems to be more consistent with the MD model in this regard. Note that, unlike T-fluctuations, the surface magnetic disorder is an extrinsic factor that, in principle, could be controlled by protecting the superconducting film from oxidation. However, in most cases, the MD is caused by the native oxide, and the corresponding broadening mechanism can be considered almost as natural as the T-fluctuation mechanism.

In summary, the analysis of transport properties in thin epitaxial TiN films reveals new microscopic mechanisms that contribute to the additional broadening of the RT curve near $T_\mathrm{c}$ compared to conventional models of superconducting fluctuations. We have successfully described the RT width on the microscopic level, proposing two mechanisms for potential inhomogeneity. Our findings suggest that these natural mechanisms could significantly affect the RT width of superconductors, which are otherwise homogeneous. 

\section{Acknowledgment} 
We are grateful to I. Burmistrov, A. Denisov, M. Feigelman, I. Gornyi, E. K\"{o}nig, A. Levchenko, A. Melnikov, D. Shovkun, and A. Shuvaev for fruitful discussions. This study was conducted as a part of RFBR, project number 19-32-60076 (development of the phenomenological model for RT broadening by magnetic impurities) and the Ministry of Science and Higher Education of the Russian Federation in the framework of the Agreement 075-11-2022-026 (sample fabrication). It was also supported by strategic project “Digital Transformation: Technologies, Effectiveness, Efficiency” of Higher School of Economics development programme granted by Ministry of science and higher education of Russia “Priority-2030” grant as a part of “Science and Universities” national project (transport measurements and the EMT model analysis) and the Basic Research Program at the HSE University (development of the model of T-fluctuations). Purdue team acknowledges support from AFOSR grant FA9550-20-1-0124 (growth of TiN films).

\section*{Appendix: Transport parameters and fluctuation regimes in TiN samples}
The crossover between 2D and 1D regimes for paraconductivity is determined by the condition $Gi_\mathrm{2D} = Gi_\mathrm{1D}$, which corresponds to a crossover width $w_\mathrm{c} \approx 15 \xi(0)\sqrt{\hbar/(e^2 R_\mathrm{s})}$. To determine the fluctuation regime in TiN samples, we use $\xi(0)^2 = -\Phi_\mathrm{0}(\dd B_\mathrm{c2}/\dd T)^{-1}/2\pi T_\mathrm{c}$, experimentally found from the temperature dependence of the second critical magnetic field $B_\mathrm{c2}(T)$, where $\Phi_\mathrm{0} = \pi \hbar/ e$ is the magnetic flux quantum. Estimates of $\xi(0)$ for 5 nm, 9 nm, and 20 nm thick TiN samples are presented in table~\ref{T1}, based on the corresponding slopes $\dd B_\mathrm{c2}/\dd T = 0.47$\,T/K, $\dd B_\mathrm{c2}/\dd T = 0.13$\,T/K, and $\dd B_\mathrm{c2}/\dd T = 0.1$\,T/K. Based on these data, we can estimate the diffusion coefficient $\mathcal{D} = 8k_\mathrm{B}T_\mathrm{c}\xi(0)^2/\pi\hbar$. The obtained values are 2.5\,cm$^2$/s, 8.5\,cm$^2$/s and 10\,cm$^2$/s for 5 nm, 9 nm, and 20 nm TiN, respectively. These values correspond to transport parameters that provide estimates of the critical width ($w_\mathrm{c}$) of 1.7 $\mu$m, 9 $\mu$m, and 12$\mu$m for the same thicknesses, respectively. Therefore, samples B1 and B2 in table~\ref{T1} are in the 2D fluctuation regime, while the rest of the samples are in the 1D fluctuation regime.


\begin{thebibliography}{10}
\expandafter\ifx\csname url\endcsname\relax
  \def\url#1{{\tt #1}}\fi
\expandafter\ifx\csname urlprefix\endcsname\relax\def\urlprefix{URL }\fi
\providecommand{\eprint}[2][]{\url{#2}}

\bibitem{Skocpol1975}
Skocpol W~J and Tinkham M 1975 {\em Rep. Prog. Phys.\/} {\bf 38} 1049--1097 \urlprefix\url{https://doi.org/10.1088/0034-4885/38/9/001}

\bibitem{Varlamov2018}
Varlamov A~A, Galda A and Glatz A 2018 {\em Rev. Mod. Phys.\/} {\bf 90}(1) 015009 \urlprefix\url{https://link.aps.org/doi/10.1103/RevModPhys.90.015009}

\bibitem{Aslamasov1968}
Aslamasov L and Larkin A 1968 {\em Phys. Lett. A\/} {\bf 26} 238--239 \urlprefix\url{https://doi.org/10.1016/0375-9601(68)90623-3}

\bibitem{Maki1968}
Maki K 1968 {\em Prog. Theor. Phys.\/} {\bf 39} 897--906 \urlprefix\url{https://doi.org/10.1143/ptp.39.897}

\bibitem{TinkhamBook}
Tinkham M 2004 {\em Introduction to {S}uperconductivity\/} (New York: Dover Publications) ISBN 9780486134727 \urlprefix\url{https://books.google.ru/books?id=VpUk3NfwDIkC}

\bibitem{Konig2021}
K\"onig E~J, Protopopov I~V, Levchenko A, Gornyi I~V and Mirlin A~D 2021 {\em Phys. Rev. B\/} {\bf 104}(10) L100507 \urlprefix\url{https://link.aps.org/doi/10.1103/PhysRevB.104.L100507}

\bibitem{Larkinlate2005}
Larkin A and Varlamov A 2005 {\em Theory of Fluctuations in Superconductors\/} (New York: Oxford University Press) \urlprefix\url{https://doi.org/10.1093/acprof:oso/9780198528159.001.0001}

\bibitem{Benfatto2009}
Benfatto L, Castellani C and Giamarchi T 2009 {\em Phys. Rev. B\/} {\bf 80}(21) 214506 \urlprefix\url{https://link.aps.org/doi/10.1103/PhysRevB.80.214506}

\bibitem{Caprara2011}
Caprara S, Grilli M, Benfatto L and Castellani C 2011 {\em Phys. Rev. B\/} {\bf 84}(1) 014514 \urlprefix\url{https://link.aps.org/doi/10.1103/PhysRevB.84.014514}

\bibitem{Sacepe2008}
Sac\'ep\'e B, Chapelier C, Baturina T~I, Vinokur V~M, Baklanov M~R and Sanquer M 2008 {\em Phys. Rev. Lett.\/} {\bf 101}(15) 157006 \urlprefix\url{https://link.aps.org/doi/10.1103/PhysRevLett.101.157006}

\bibitem{Hortensius2013}
Hortensius H~L, Driessen E~F~C and Klapwijk T~M 2013 {\em IEEE Transactions on Applied Superconductivity\/} {\bf 23}(3) 2200705 \urlprefix\url{https://ieeexplore.ieee.org/abstract/document/6407809/}

\bibitem{Venditti2019}
Venditti G, Biscaras J, Hurand S, Bergeal N, Lesueur J, Dogra A, Budhani R~C, Mondal M, Jesudasan J, Raychaudhuri P, Caprara S and Benfatto L 2019 {\em Phys. Rev. B\/} {\bf 100}(6) 064506 \urlprefix\url{https://link.aps.org/doi/10.1103/PhysRevB.100.064506}

\bibitem{Mondal2011}
Mondal M, Kumar S, Chand M, Kamlapure A, Saraswat G, Seibold G, Benfatto L and Raychaudhuri P 2011 {\em Phys. Rev. Lett.\/} {\bf 107}(21) 217003 \urlprefix\url{https://link.aps.org/doi/10.1103/PhysRevLett.107.217003}

\bibitem{Saveskul}
Saveskul N, Titova N, Baeva E, Semenov A, Lubenchenko A, Saha S, Reddy H, Bogdanov S, Marinero E, Shalaev V, Boltasseva A, Khrapai V, Kardakova A and Goltsman G 2019 {\em Phys. Rev. Applied\/} {\bf 12}(5) 054001 \urlprefix\url{https://link.aps.org/doi/10.1103/PhysRevApplied.12.054001}

\bibitem{first}
Baeva E~M, Kolbatova A~I, Titova N~A, Saha S, Boltasseva A, Bogdanov S, Shalaev V~M, Semenov A~V, Levchenko A, Goltsman G~N and Khrapai V~S 2024 {\em Phys. Rev. B\/} {\bf 110}(10) 104519 \urlprefix\url{https://link.aps.org/doi/10.1103/PhysRevB.110.104519}

\bibitem{Suppl_Data}
See Supplemental Material for details on: (i) images of devices of different configurations; (ii) estimates of the contact resistance; (iii) the experimental data of the resistive transition at different excitation currents for one representative sample; (iv) - (v) the temperature dependences of the normalized resistance $R/R_n$ for all studied samples in comparison with the conventional models of superconducting fluctuations and in comparison to the effective medium theory (EMT) model.

\bibitem{Thompson1970}
Thompson R~S 1970 {\em Phys. Rev. B\/} {\bf 1}(1) 327--333 \urlprefix\url{https://link.aps.org/doi/10.1103/PhysRevB.1.327}

\bibitem{Ioffe1993}
Ioffe L~B, Larkin A~I, Varlamov A~A and Yu L 1993 {\em Phys. Rev. B\/} {\bf 47}(14) 8936--8941 \urlprefix\url{https://link.aps.org/doi/10.1103/PhysRevB.47.8936}

\bibitem{Johnson_1978}
Johnson W~L, Tsuei C~C and Chaudhari P 1978 {\em Phys. Rev. B\/} {\bf 17}(7) 2884--2891 \urlprefix\url{https://link.aps.org/doi/10.1103/PhysRevB.17.2884}

\bibitem{Hopf1991}
Hopfeng\"artner R, Hensel B and Saemann-Ischenko G 1991 {\em Phys. Rev. B\/} {\bf 44}(2) 741--749 \urlprefix\url{https://link.aps.org/doi/10.1103/PhysRevB.44.741}

\bibitem{Gauzzi1995}
Gauzzi A and Pavuna D 1995 {\em Phys. Rev. B\/} {\bf 51}(21) 15420--15428 \urlprefix\url{https://link.aps.org/doi/10.1103/PhysRevB.51.15420}

\bibitem{AG61}
Abrikosov A and Gorkov L 1961 {\em Sov. Phys. JEPT\/} {\bf 12} 1243

\bibitem{goldman1984percolation}
Goldman A 1984 {\em Percolation, Localization, and Superconductivity\/} (New York: Plenum Press) ISBN 978-1461593966 \urlprefix\url{https://doi.org/10.1007/978-1-4615-9394-2}

\bibitem{Halperin1979}
Halperin B~I and Nelson D~R 1979 {\em J. Low Temp. Phys.\/} {\bf 36} 599--616 ISSN 1573-7357 \urlprefix\url{https://doi.org/10.1007/BF00116988}

\bibitem{Konig2015}
K\"onig E~J, Levchenko A, Protopopov I~V, Gornyi I~V, Burmistrov I~S and Mirlin A~D 2015 {\em Phys. Rev. B\/} {\bf 92}(21) 214503 \urlprefix\url{https://link.aps.org/doi/10.1103/PhysRevB.92.214503}

\bibitem{Gantmakher_2010}
Gantmakher V~F and Dolgopolov V~T 2010 {\em Physics-Uspekhi\/} {\bf 53} 1--49 \urlprefix\url{https://doi.org/10.3367/ufne.0180.201001a.0003}

\bibitem{Baity2016}
Baity P~G, Shi X, Shi Z, Benfatto L and Popovi\ifmmode~\acute{c}\else \'{c}\fi{} D 2016 {\em Phys. Rev. B\/} {\bf 93}(2) 024519 \urlprefix\url{https://link.aps.org/doi/10.1103/PhysRevB.93.024519}

\bibitem{Kittel}
Kittel C 2012 {\em Introduction to Solid State Physics\/} 8th ed (New York: Wiley) \urlprefix\url{https://www.wiley.com/en-us/Introduction+to+Solid+State+Physics%2C+8th+Edition-p-9780471415268}

\bibitem{Proslier2008}
Proslier T, Zasadzinski J~F, Cooley L, Antoine C, Moore J, Norem J, Pellin M and Gray K~E 2008 {\em Applied Physics Letters\/} {\bf 92} 212505 \urlprefix\url{https://doi.org/10.1063/1.2913764}

\bibitem{Sendelbach2008}
Sendelbach S, Hover D, Kittel A, M\"uck M, Martinis J~M and McDermott R 2008 {\em Phys. Rev. Lett.\/} {\bf 100}(22) 227006 \urlprefix\url{https://link.aps.org/doi/10.1103/PhysRevLett.100.227006}

\bibitem{deGraaf2020}
de~Graaf S~E, Faoro L, Ioffe L~B, Mahashabde S, Burnett J~J, Lindström T, Kubatkin S~E, Danilov A~V and Tzalenchuk A~Y 2020 {\em Science Advances\/} {\bf 6} eabc5055 \urlprefix\url{https://www.science.org/doi/abs/10.1126/sciadv.abc5055}

\bibitem{Fang2020}
Yang F, Gozlinski T, Storbeck T, Gr\"unhaupt L, Pop I~M and Wulfhekel W 2020 {\em Phys. Rev. B\/} {\bf 102}(10) 104502 \urlprefix\url{https://link.aps.org/doi/10.1103/PhysRevB.102.104502}

\bibitem{tamir2021direct}
Tamir I, Trahms M, Gorniaczyk F, von Oppen F, Shahar D and Franke K~J 2022 {\em Phys. Rev. B\/} {\bf 105}(14) L140505 \urlprefix\url{https://link.aps.org/doi/10.1103/PhysRevB.105.L140505}

\bibitem{Kuzmiak2022}
Kuzmiak M, Kop{\v{c}}{\'i}k M, Ko{\v{s}}uth F, Va{\v{n}}o V, Szab{\'o} P, Latyshev V, Komanick{\'y} V and Samuely P 2022 {\em Journal of Superconductivity and Novel Magnetism\/} {\bf 35} 1775--1780 ISSN 1557-1947 \urlprefix\url{https://doi.org/10.1007/s10948-022-06197-6}

\bibitem{Koshelev2020}
Koshelev A~E 2020 {\em Phys. Rev. B\/} {\bf 102}(5) 054505 \urlprefix\url{https://link.aps.org/doi/10.1103/PhysRevB.102.054505}

\end{thebibliography}
\providecommand{\newblock}{}

\newpage
\appendix
\end{document}


\title[SUPPLEMENTARY MATERIAL]{SUPPLEMENTARY MATERIAL for \\ Natural width of the superconducting transition in epitaxial TiN films}

\author{E.M. Baeva$^{1,2}$, A.I. Kolbatova$^{1}$, N.A. Titova$^{1}$, S. Saha$^{3}$,\\ A. Boltasseva$^{3}$, S. Bogdanov$^{4,5,6}$, V.M.  Shalaev$^{3}$,\\ A.V. Semenov$^{2}$, G.N. Goltsman$^{1,2}$, and V.S. Khrapai$^{2}$}

\address{$^1$ Moscow Pedagogical State University, 119435 Moscow, Russia}
\address{$^2$ HSE University, 101000 Moscow, Russia}
\address{$^3$ Birck Nanotechnology Center and Elmore Family School of Electrical and Computer Engineering, Purdue University, West Lafayette, IN 47907, USA}
\address{$^4$ Department of Electrical and Computer Engineering, University of Illinois at Urbana-Champaign, Urbana, IL 61801, USA}
\address{$^5$ Holonyak Micro and Nanotechnology Lab, University of Illinois at Urbana-Champaign, Urbana, IL 61801, USA}
\address{$^6$ Illinois Quantum Information Science and Technology Center, University of Illinois at Urbana-Champaign, Urbana, IL 61801, USA}

\begin{abstract}
In this supplementary material, we present (i) images of devices with different configurations; (ii) estimates of contact resistance; (iii) experimental data on the resistive transition for different excitation currents for a representative sample; (iv) temperature dependencies of normalized resistance $R/R_n$ for all the studied samples comparison with conventional models of superconducting fluctuations, and (v) comparison with the effective medium theory (EMT).
\end{abstract}

\maketitle

\begin{figure}[h!]
\centering
\includegraphics[scale=1]{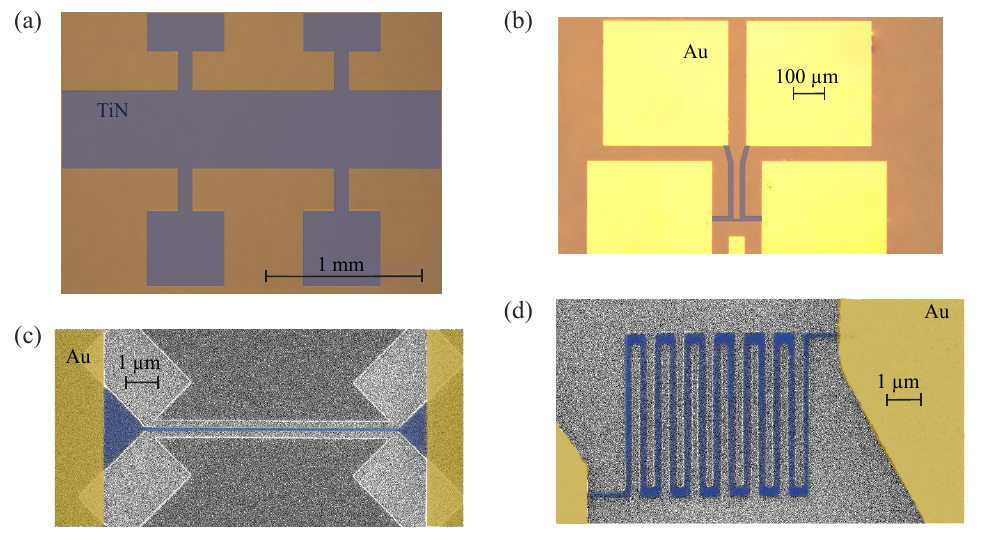}
\caption{Optical and SEM (scanning electron microscopy) images of representative configurations of the studied samples: (a) a four-probe Hall bar configuration (samples A4 and B1 in table~1 in the main manuscript); (b) a four-point contact configuration (B2, B3, B4, and B5); (c) a two-contact bridge configuration (A1-A3 and B6-B9); (d) a meander configuration (A3 and C1). \label{figure_num_1}}
\end{figure}

\begin{figure}[h!]
\centering
\includegraphics[scale=1]{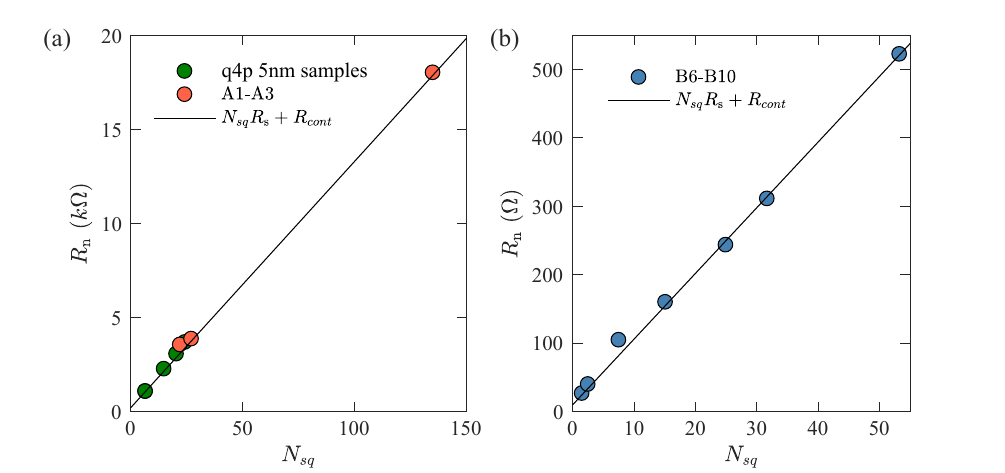}
\caption{Experimental approach for evaluating the contact resistance: The normal-state resistance $R_\mathrm{n}$ as a function of the number of squares $N_\mathrm{sq} = L/w$ for (a) 5 nm and (b) 9 nm thick TiN samples. In (a), red symbols represent $R_\mathrm{n}$ measured at 300\,K in a quasi-four-probe configuration (q4p) for samples A1-A3. To accurately determine the contact resistance $R_\mathrm{cont}$ for 5 nm thick samples, we added data from other 5 nm samples (green symbols) that were prepared using the same process as A1-A3. In (b), data corresponds to $R_\mathrm{n}$ measured at 6\,K in q4p for samples B6-B10. Black solid lines represent a linear fit $R_\mathrm{n} = N_\mathrm{sq}R_\mathrm{s} + R_\mathrm{cont}$, with fitting value $R_\mathrm{cont} \simeq 200$\,$\Omega$ for 5 nm and $R_\mathrm{cont} \simeq 10$\,$\Omega$ for 9 nm.\label{figure_num_2}}
\end{figure}

\begin{figure}[h!]
\centering
\includegraphics[scale=1]{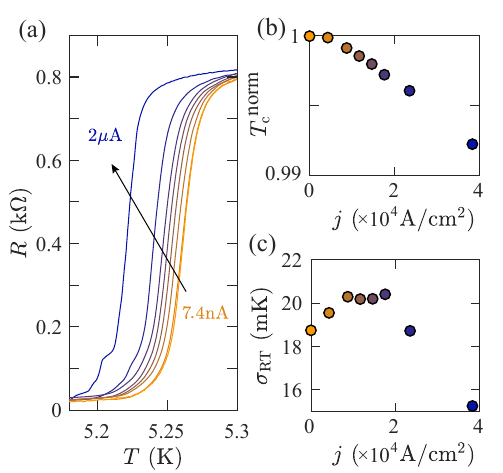}
\caption{The resistive transition at different excitation currents: (a) The resistive transition of a 20 nm thick sample (C1), recorded at zero magnetic field, at different bias currents. As the bias current increases from 7\,nA to 2\,$\mu$A, the resistive transition shifts downward. (b) The normalized critical temperature $T_c/T_{c0}$ as a function of bias current density $j$. $T_c$ is determined at the transition midpoint, where $R = R_n/2$. (c) The RT width $\sigma_{RT}$ versus $j$. The experimental curves show that $T_c$ slightly decreases and $\sigma_{RT}$ non-monotonously changes with increasing the excitation current. Maximum changes in $T_c$ and $\sigma_{RT}$, which are observed at given bias currents, are about 1\% and 22\%, respectively. 
\label{figure_num_2}}
\end{figure}

\begin{figure}[h!]
\centering
\includegraphics[scale=1]{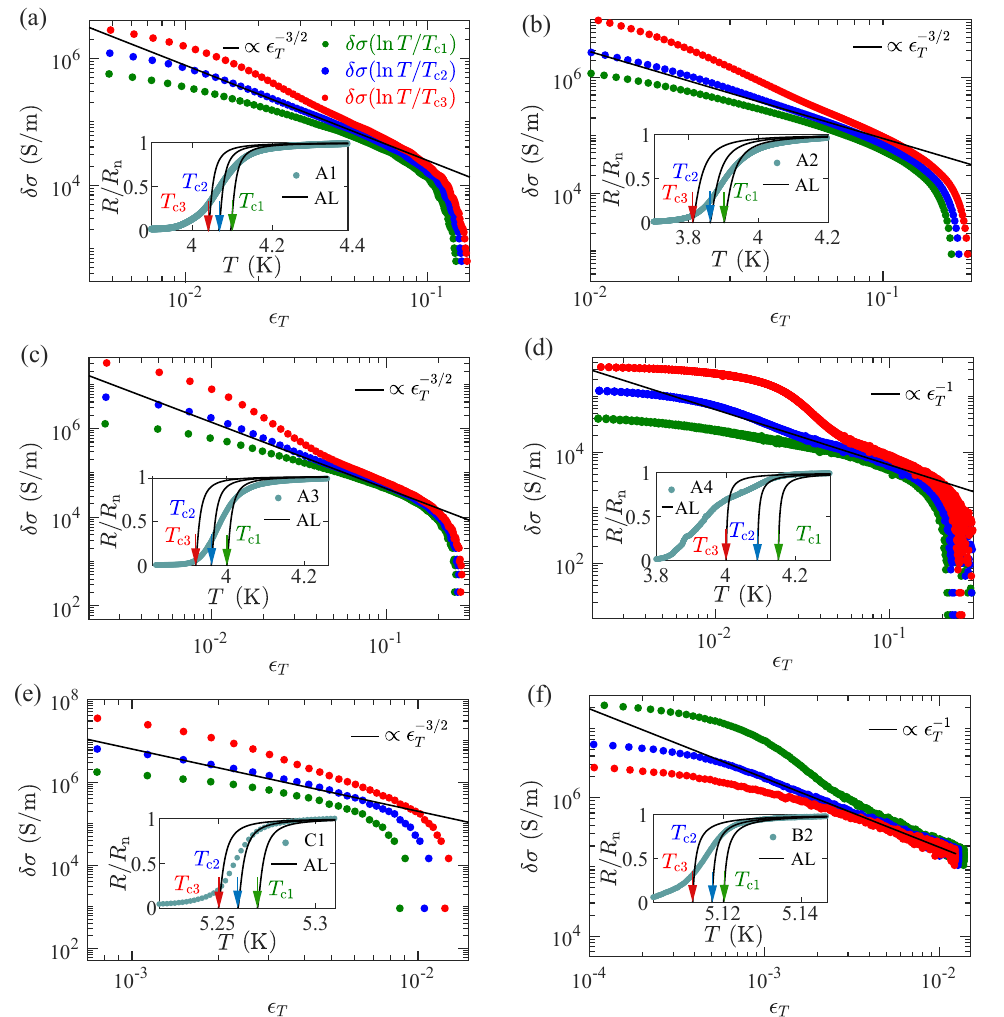}
\caption{The resistive transition and conventional superconducting fluctuations are shown for all TiN samples studied in this work. The main body of the figure shows the fluctuation conductivity, $\delta\sigma = \rho(T)^{-1}-\rho_\mathrm{n}^{-1}$, versus the reduced temperature, $\epsilon_\mathrm{T} = \ln(T/T_\mathrm{c0})$, computed at three different values of $T_\mathrm{c0}$, indicated by arrows in the inset. The data are plotted on a double logarithmic scale. The solid lines in the figures (a)-(c), (e) represent the 1D regime of the Aslamazov-Larkin (AL) model ($\propto\epsilon_\mathrm{T}^{-3/2}$), (d) and (f) the 2D AL regime ($\propto\epsilon_\mathrm{T}^{-1}$). In the inset, the normalized resistance, $R/R_\mathrm{n}$, is plotted versus $T$. The solid black lines represent predictions of the AL model computed at three different values of $T_\mathrm{c0}$. \label{figure_num_3}}
\end{figure}

\begin{figure}[h!]
\centering
\includegraphics[scale=1]{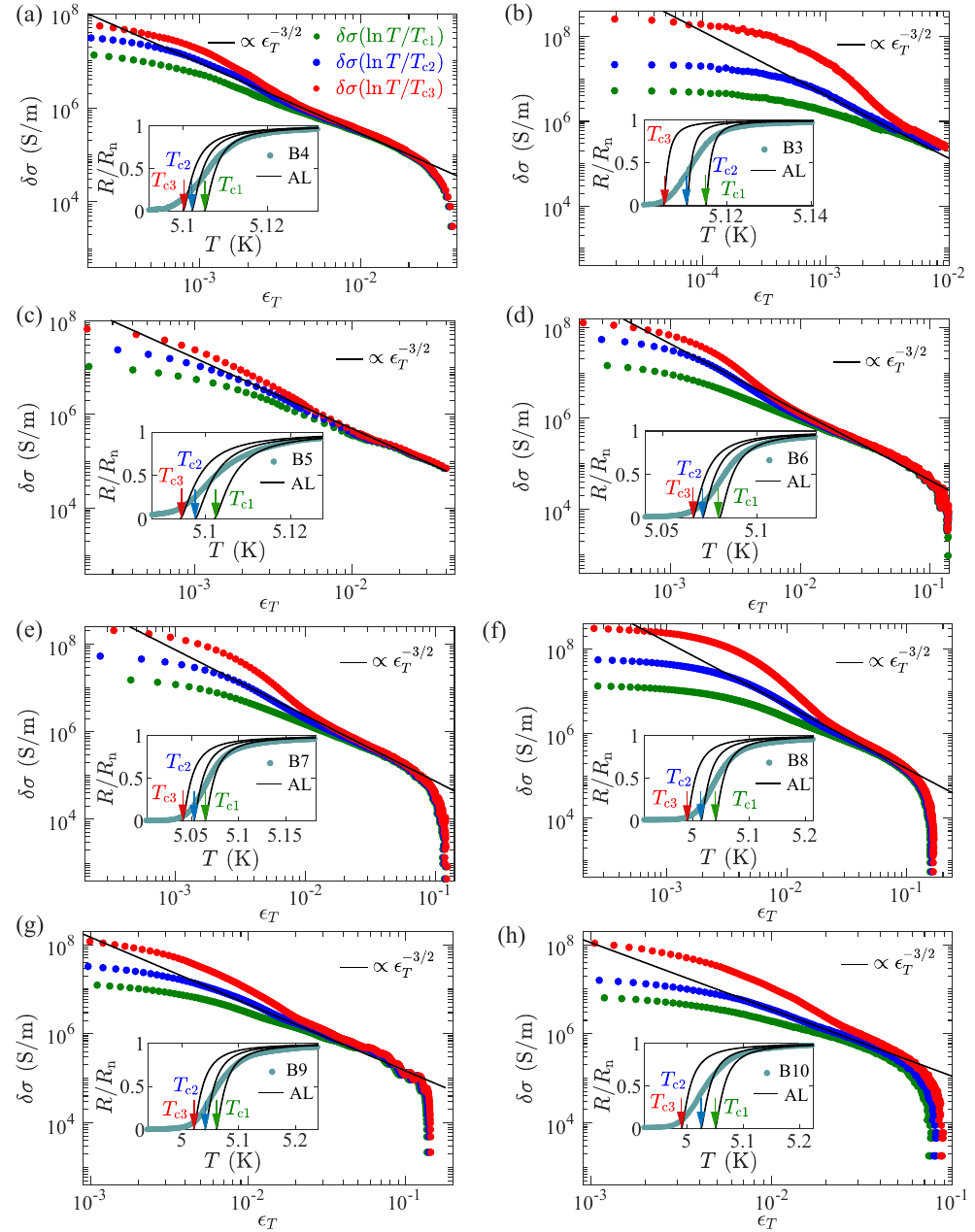}
\caption{The resistive transition and conventional superconducting fluctuations are shown for all TiN samples studied in this work. The main body of the figure shows the fluctuation conductivity, $\delta\sigma = \rho(T)^{-1}-\rho_\mathrm{n}^{-1}$, versus the reduced temperature, $\epsilon_\mathrm{T} = \ln(T/T_\mathrm{c0})$, computed at three different values of $T_\mathrm{c0}$, indicated by arrows in the inset. The data are plotted on a double logarithmic scale. The solid lines in the figures (a)-(c), (e) represent the 1D regime of the Aslamazov-Larkin (AL) model ($\propto\epsilon_\mathrm{T}^{-3/2}$), (d) and (f) the 2D AL regime ($\propto\epsilon_\mathrm{T}^{-1}$). In the inset, the normalized resistance, $R/R_\mathrm{n}$, is plotted versus $T$. The solid black lines represent predictions of the AL model computed at three different values of $T_\mathrm{c0}$. \label{figure_num_4}}
\end{figure} 

\begin{figure}[h!]
\centering
\includegraphics[scale=1]{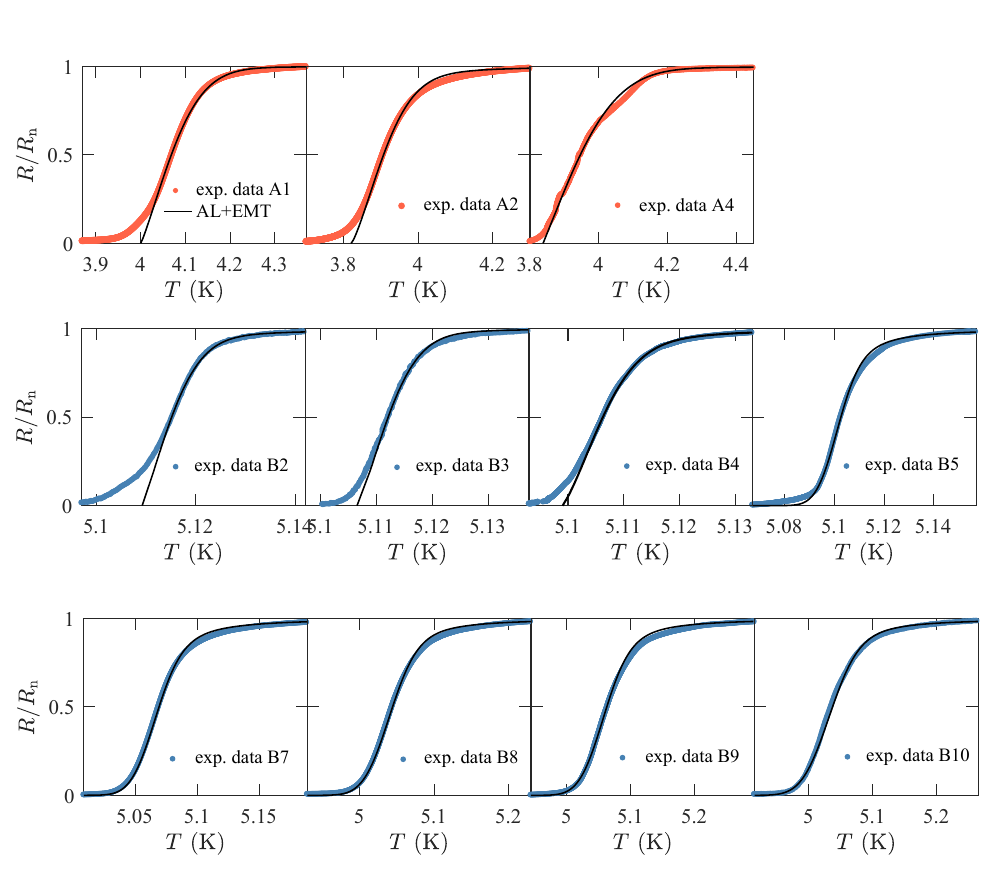}
\caption{The temperature dependence of the normalized resistance, $R/R_\mathrm{n}$, for the studied TiN samples and the corresponding fits using the EMT model. The experimental data is represented by symbols, while the results of the fitting with the EMT model are shown with black solid lines. Within the framework of the magnetic disorder (MD) scenario, samples A1-A4 and B2-B4, are in 2D regime, while the samples B7-B10, C1 are in 1D regime.  \label{figure_num_5}}
\end{figure} 

%

\newpage
\appendix